**A study of the magnetotransport properties of the graphene (III. Bilayer)**

M. A. Hidalgo [a], R. Cangas [b]

[a] Departamento de Física y Matemáticas, Universidad de Alcalá, Alcalá de Henares (Madrid), Spain

[b] Departamento de Física, Escuela Técnica de Ingeniería Industrial, Universidad Politécnica de Madrid, Madrid, Spain

Correspondence and request for materials should be addressed to miguel.hidalgo@uah.es

(*) *All the references by the authors can be downloaded from their "Researchgate" website*

**Abstract**

The present paper corresponds to the third work of the author related to the magnetotransport properties concerning on the graphene systems. In the first one the integer quantum Hall effect in the monolayer graphene, (MG), MGIQHE, was analysed, (Hidalgo, 2014). The subject of the second one was the understanding of the fractional quantum Hall also in the MG, (MGFQHE), (Hidalgo, 2015), basing us in our model on the fractional quantum Hall effect in semiconductor quantum systems, Hidalgo (2013). The aim of the present work is the analysis of the integer quantum Hall effect observed in bilayer graphene, (BG), BGIQHE, as a function of both, the gate voltage and the magnetic field. Our approach is a single electron approach, firstly developed for the study of the integer quantum Hall (IQHE) and Shubnikov-de Haas (SdH) effects of a two-dimensional electron gas (2DEG) in any semiconductor quantum well (QW), (Hidalgo, 1995, 1998, 2007). The extension of this model for the BG reproduces its main observed features for the IQHE and the SdH, and as a function of both, the gate voltage and the magnetic field, in particular the plateaux at the values $\pm 4n$, with $n$=0,1,2... Therefore, the approach integrates the quantum Hall effects (integer and fractional) observed in both, QW and graphene, in the same physical frame.

**Introduction**

The integer and fractional quantum Hall effects, (IQHE and FQHE, respectively) in QW were important in solid state physics, not only for their intrinsic interest from a fundamental point of view, but for their potential applications. Nowadays the observation of similar phenomena in graphene systems (monolayer and bilayer) has reactivated the interest on these magnetotransport properties, associated with any two-dimensional electron system (2DES) embedded in a 3D system. In fact, the features of all of them are identical: minima or even zeroes in the longitudinal resistance, (SdH), and simultaneous with them well-defined plateaux in the Hall resistance at integer or fractional values of the fundamental Hall resistance $R_H = h/e^2$).

It is well-known that the IQHE measured in any 2DES in a QW shows plateaux following the integer series $2n$ (i.e., 2, 4, 6, 8…) when there is spin degeneration, or $n$ (i.e., 1, 2, 3, 4, 5…) when that spin degeneration is broken (Hidalgo, 1995). Several years ago we already developed an alternative global approach to the IQHE and FQHE in QW in the context of the single particle approximation, (Hidalgo, 1995, 1998, 2007, 2013).

More recently, similar effects have also been observed in MG, where the plateaux appear at values the $\pm 2(2n+1)$, with $n=0,1,2…$, and the BG where the sequence of the plateaux is given by the general expression $\pm 4n$, i.e., 0, ±4, ±8, ±12, ±16, ±20, ±24… It is thought that the existence of an external bias induced band gap in BG, differing from MG, and that the origin of the unconventional BGIQHE lies in the coupling between the two graphene layers, transforming massless Dirac fermions, characteristic of single layer graphene into a new type of chiral quasiparticle. However, in the present paper we show how a single electron approach, based in first principles, is capable to explain the quantum Hall effects observed in MG and BG, having the same origin as the ones observed in QW.

In our paper Hidalgo 2014 we extended our magnetotransport model developed for understanding the IQHE in QW, (Hidalgo 1998, 1997 and 2007), to the magnetotransport in MG. We showed this extension was capable to understand all the physical properties associated with. Therefore, below we describe how is used the same model to understand those phenomena in BG.

The structure of the paper is the following: in the next section we determine the density of states for the 2DES in BG, crucial for obtaining the magnetotransport magnitudes, which we detail in the third section. In the fourth one we describe the results of the model, showing that it provides all the aspects of those phenomena. Finally, we have added a summary and discussion section.

**The density of states for the monolayer graphene**

BG is a two layer of carbon atoms arranged in a hexagonal lattice with two carbon atoms per unit cell, (Goerbig, 2011), and with a fixed interlayer separation of 0.3 nm, similar to graphite; in fact two graphene monolayers that are weakly coupled by interlayer carbon hopping. The spin and valley degeneracy in BG is equal to 4.

The BG system is intermediate between the MG and the bulk graphite. A detailed description of the bands structure in both, MG and BG, is found elsewhere (Goerbig 2011, Das Sarma 2011). But here it is important to highlight that the main difference between both is, as theoretical calculations predict, the existence of a gap in the BG, (McCann 2006, and McCann and Falko 2006). Then, in comparison with MG, (Hidalgo 2014), the main consequence for us of this fact is that there is no any connection between both bands, what implies that the dynamical spaces under the application of **B** associated with each sub-band, $\varepsilon_{VB}^{\mathbf{B}}$ and $\varepsilon_{CB}^{\mathbf{B}}$, are independent each other, unlike in MG. (Hidalgo 2014).

Therefore, to determine the energy spectrum of electrons in BG, we first calculate the quantized electron states under that condition. We will take into account the symmetric gauge with the magnetic field perpendicular to the 2DES of graphene, $\mathbf{B} = (0,0,B)$.

Thus, as we have mentioned above, the contribution of each sub-band in BG is independent, and the energy states are then given by the following expression

$$E_n = \left(n + \frac{1}{2}\right)\hbar\omega_0 = \left(n + \frac{1}{2}\right)E_0 \qquad (1)$$

n=0, 1, 2…, $\omega_0 = eB/m^*$ being the fundamental angular frequency, and $E_0 = \hbar\omega_0$.

Once we have the expression for the energy spectrum, the magnitude necessary to determine any physical property is the corresponding density of states. From equation (1), using the Possion summation formula, (Schoenberg, 1984; Hidalgo 1995, 1998, 2007), taking into account the spin-orbit interaction, we obtain the expression for the density of states of each band, given by

$$g(E) = g_0\left\{1 + 2\sum_{p=1}^{\infty} A_{S,p} A_{\Gamma,p} \cos\left[2\pi p\left(\frac{E}{\hbar\omega_0} - \frac{1}{2}\right)\right]\right\} \qquad (2)$$

where p is the summation index corresponding to the p harmonic, and $g_0$ the two-dimensional density of states in absence of magnetic field. Looking at the expression of the density of states we found for the MG, the difference with equation (2) is the term ½ appearing now in the argument of cosine.

$A_{S,p} = \cos\left(\pi p \frac{g^*}{2} \frac{m^*}{m}\right)$ is the term associated with the spin and spin-orbit coupling, where $g^*$ is the generalized gyromagnetic factor, -that we assume here to be the same for both bands and having a value of 2, in correspondence with the experimental measurements, (Zhang et al. 2006).

$$A_{\Gamma,p} = exp\left\{-\frac{2\pi^2 p^2 \Gamma^2}{\hbar^2 \omega_0^2}\right\}$$ is related to the width of the energy levels, determined directly by the factor $\Gamma$. This term is due to the interaction of the electron with defects and ionized impurities in the system. In all bellow, for sake of the simplicity, we assume constant gaussian width for every energy level.

The next step in the development of the model will be to include the two valley degeneracy, corresponding to the K and K' points, characteristic of the MG and BG, and whose corresponding density of states we suppose to be identical as a first approach. Hence, the total density of states of the 2DES in the BG will be expressed by the equation

$$g^{total}(E) = g^{K}(E) + g^{K'}(E) \tag{3}$$

where the indexes refer to the density of states of each DP, given by equation (2).

**The model for both magnetoconductivities**

As we have mentioned above, our approach to the magnetotrasport properties of the BG will be based on the same steps given in our previous work for the IQHE and FQHE in MG, and based in a single electron approximation, (Hidalgo 2014).

We will initially consider the DP K. Taking the magnetotransport expressions for the 2DES in the semiconductor QW developed in the references by Hidalgo (1995, 1998, 2007), we have for the diagonal magnetoconductivity

$$\sigma_{xx}^{K} = \frac{1}{\omega_0 \tau} \frac{en^{K}}{B} \frac{g^{K}(E_F)}{g_0} \tag{4}$$

where $g(E_F)$ is the density of states at the Fermi level as obtained from equation (2), and $\tau$ the relaxation time of any electrons in the corresponding Dirac point and band. On the other hand, for the Hall magnetoconductivity at high magnetic fields we have

$$\sigma_{xy}^{K} = -\frac{en^{K}}{B} \tag{5}$$

being $n^{K}$ the electron density at K, as easily obtained from the equation (2)

$$n = n_0 + \frac{2eB}{h}\sum_{p=1}^{\infty}\frac{1}{\pi p}A_{S,p}A_{\Gamma,p}A_{T,p}sen\left[2\pi p\left(\frac{E_F}{\hbar\omega_0}-\frac{1}{2}\right)\right] = n_0 + \delta n \tag{6}$$

with $n_0$ the density of electrons at zero magnetic field (or zero gate voltage); and $\delta n$ the variation in the electron density as a consequence of the quantized density of states.

The effect of the temperature, which ultimately determines the occupation of every state, is incorporated in the term $A_{T,p} = z/senh(z)$, where $z = 2\pi^2 pkT/\hbar\omega_0$, $k$ being the Boltzmann constant and $T$ the corresponding temperature.

Equations (4) and (5) are the magnetoconductivities expressions of the model not only for K but also for the point K'. Now, because we are assuming both are equivalent we can add the corresponding magnetoconductivities, obtaining for the Hall one

$$\sigma_{xy}^{total} = \sigma_{xy}^{K} + \sigma_{xy}^{K'} = -\frac{en^{total}}{B} \tag{7}$$

where $n^{total}$ is the total density of electrons, i.e. $n^{total} = n^{K} + n^{K'} = 2n^{K}$. And for the diagonal magnetoconductivity,

$$\sigma_{xx}^{total} = \sigma_{xx}^{K} + \sigma_{xx}^{K'} = \frac{1}{\omega_0\tau}\frac{en^{total}}{B}\frac{g^{total}(E_F)}{g_0} \tag{8}$$

$\tau$ is the relaxation time of the electrons in each DP that here we assume as a first approach to be the same.

From equations (7) and (8) is immediate to calculate the symmetric magnetoresistivity tensor, $[\rho] = [\sigma]^{-1}$, whose terms are determined by the expressions

$$\rho_{xx} = \rho_{yy} = \frac{\sigma_{xx}}{\sigma_{xx}^2 + \sigma_{xy}^2} \tag{9}$$

$$\rho_{xy} = -\rho_{yx} = -\frac{\sigma_{xy}}{\sigma_{xx}^2 + \sigma_{xy}^2} \tag{10}$$

**Simulations**

This section is devoted to the analysis of the results of the model and its comparison with some experimental measurements. We show how the model performs as a function of the gate voltage and the magnetic field and, also, how is the evolution of the magnetotransport magnitudes with temperature.

From the evolution with temperature of the maxima of the SdH oscillations, it is possible to determine the effective mass for carriers in the BG, finding that has a value of the order of $m^*=(0,03-0,05)m_0$ ($m_0$ being the free electron mass), (Zhao et al. 2010).

1) GQHE as a function of the gate voltage:

For testing the model, firstly we simulate the experimental measurements in BG as a function of the gate voltage. Calculating the variation of the electron density in the 2DES as a function of the applied gate voltage, $V_g$, and assuming a linear relation between both magnitudes, we can write the Fermi level in equation (8) as

$$E_F = eV_g \tag{11}$$

In Figure 1 we show the results obtained with the model for the Hall magnetoconductivity, (a), and the Hall and diagonal magnetoresistivities, (b), at the gate voltage interval between -7 and 7 V. To achieve the simulation we have assumed a temperature of 1.6 K, a gyromagnetic factor of 2; and a fixed magnetic field $B$=25 T. The effective mass considered is $m^*=0,05m_0$, (Zhao et al. 2010). Additionally, we have assumed for the simulation a constant Gaussian width with $\Gamma$=0.06 eV and a relaxation time $\tau$=1 ps. As it is clearly seen in the figures, the model reproduces accurately the experimental results, appearing the plateaux in the Hall magnetoconductivity at the values given by

$$\sigma_{xy} = \pm 4n \frac{e^2}{h} \qquad (12)$$

in correspondence with the observed plateaux series 0, ±4, ±8, ±12, ±16, ±20, ±24…

It is also seen a common feature of all quantum Hall effects: the gate voltage intervals where the SdH oscillations are minima match with the plateaux ones.

For the sake of completeness, in Figure 2 it is shown the Hall magnetoconductivity as a function also of the gate voltage but for different magnetic fields (a) 15 T, (b) 25 T (the rest of the conditions assumed are the same as in Figure 1).

    2) GQHE as a function of the magnetic field:

Although not many measurements are found in literature related to the magnetotransport properties of BG as a function of the magnetic field, the model allows us to analyse them. In these cases the Fermi level will be fixed and given by

$$E_F = \frac{\pi \hbar^2 n_0}{m^*} \qquad (13)$$

where $n_0$ is the electron density of the 2DES at zero magnetic field, (directly obtained from the Hall measurements at low magnetic fields). Determining the magnetic field values of the experimental maxima of the SdH oscillations, we can easily see that they match with the magnetic fields given by the expression

$$B_n \approx \frac{m^* E_F}{\hbar e \left(n + \frac{1}{2}\right)} \qquad (14)$$

In fact, it shows that $n$ is related to the plateaux index $4n$. (The same equation (14) can be used to predict the position of the maxima of the SdH oscillations in the experiments of the IQHE in QW, once we know the electron density at zero magnetic field. Moreover, a similar expression between the quantized levels and the plateaux is found in the FQHE, (Hidalgo 2013)).

In Figure 3 shows the simulations obtained with the model for the Hall magnetoconductivity, (a), and diagonal magnetoresistivities, (b), as a function of the magnetic field, and for a sample at a temperature of 1.6 K; with an electron density $n_0=3\times10^{16}$ m$^{-2}$; a gyromagnetic factor of 2; a constant Gaussian width for the energy levels with $\Gamma=0.025$ eV; a relaxation time $\tau=1$ ps; and an effective mass of $m^*=0.05m_0$.

The plateaux appear at the right places in the entire magnetic field interval.

**Summary and Discussion**

We have presented the extension of the model previously developed for the MG, (Hidalgo, 2014), a single electron approach whose origin is the model already built for the IQHE in QW, (Hidalgo, 1995, 1998 and 2007). It reproduces all the observed characteristics of the magnetotransport properties in BG; integrating this phenomenon in the same framework as the IQHE and the FQHE: the same first physical principles and similar basic hypothesis and assumptions.

With the simulations made with the model, Figures (1)-(3), we show that the model is capable of explaining in a simple way the series of the Hall plateaux observed experimentally and, moreover, their appearance even at high temperatures. Hence, we hope that it can be a useful tool in the understanding of the conduction mechanisms and magnetotransport properties.

The appearance of plateaux at values 0, 1 and 4 in MG, and the sequence appearing in the works by Zhao et al (2010), will be discussed in a forthcoming paper in light of the model.

**References**

1. Hidalgo, M. A., *PhD Thesis*. Editorial de la Universidad Complutense de Madrid (1995) (*)

2. Hidalgo, M. A., **43-44**, 453 (1998) (*)

3. Hidalgo, M. A., Cangas, R., *Arxiv*: **0707.4371** (2007) (*)

4. Hidalgo, M. A., *Arxiv*: **1310.1787** (2013) (*)

5. Hidalgo, M. A., *Arxiv*: **1404.5537**. doi: 10.13140/2.1.2006.5923 (2014) (*)

6. Hidalgo, M. A., *Arxiv*: **1507.05023**. doi: 10.13140/RG 213737.9684 (2015) (*)

7. Novoselov, K. S. et al., *Nature* **doi:10.1038/nphys245** (2006)

8. Castro Neto, A. H., *Rev. Mod. Phys.* **81**, 109 (2009)

9. Goerbig, M. O., *Rev. Mod. Phys.* **83**, 1193 (2011)

10. Das Sarma, S. et al., *Rev. Mod. Phys.* **83**, 407 (2011)

11. Janssen, T. J. B. M. et al., *Rep. Prog. Phys.* **76**, 104501 (2013)

12. McCann, E., *Phys. Rev.* B**74**, 161403(R) (2006)

13. McCann, E., et al. *Phys. Rev. Lett.* **96**, 086805 (2006)

14. Schoenberg, D., Magnetic oscillations in metals. Cambridge University Press (1984)

15. Zhang, Y. et al., *Phys. Rev. Lett.* **96**, 136806 (2006)

16. Zhao, Y. et al. *Phys. Rev. Lett.* **104**, 066801 (2010)

(*) *All the references by the authors can be downloaded from their "Researchgate" website*

**Figure legends:**

**Figure 1: Simulation with the model of the Hall magnetoconductivity, (a), and the Hall and diagonal magnetoresistivities, (b), of the bilayer graphene as a function of the gate voltage**: The gate voltage interval is -7 to 7 V. The conditions of the simulation are: a temperature of 1.6 K; a magnetic field of $B$=25 T [28], and a gyromagnetic factor of 2. The effective mass assumed is $m^*$=0,05$m_0$. Besides, we have considered a constant Gaussian width with $\Gamma$=0.06 eV and a relaxation time $\tau$=1 ps. In figure (a) the integer numbers over the reference lines label the different plateaux.

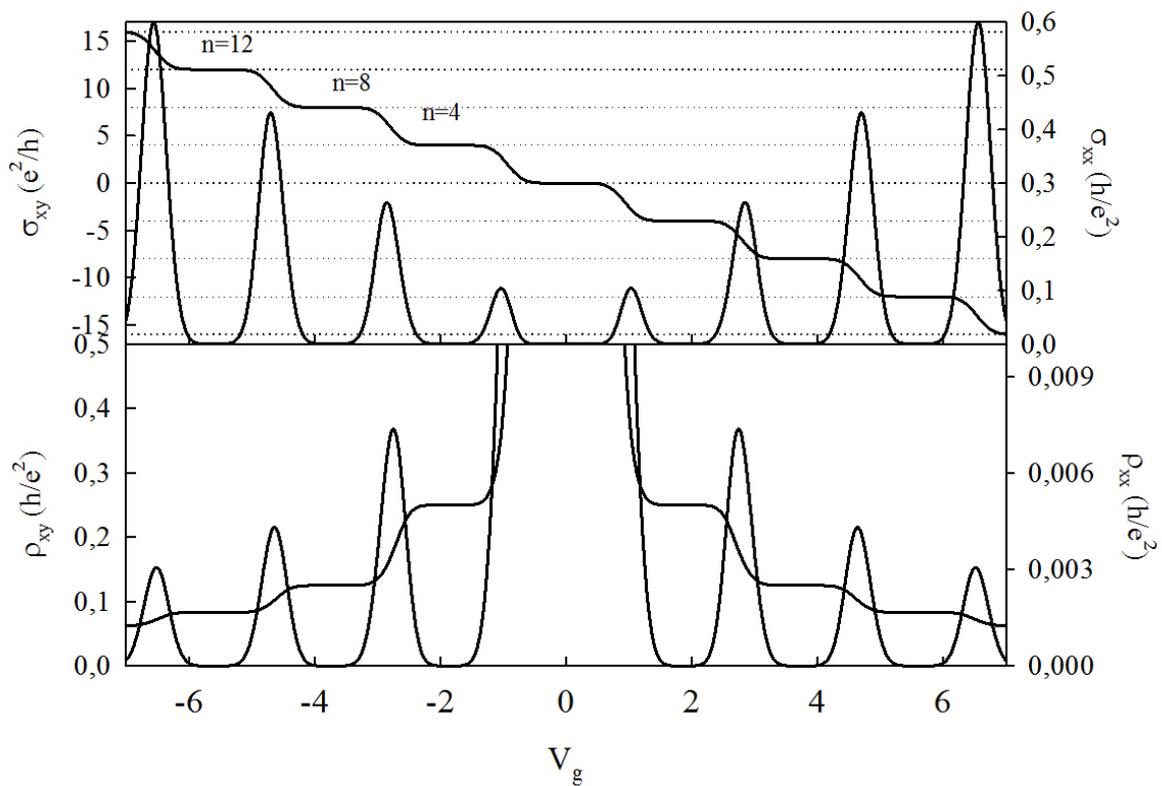

**Figure 2: Simulation with the model of the Hall magnetoconductivity for the bilayer graphene as a function of the gate voltage for different magnetic field values**: The magnetic fields fixed are (a) 15 T and (b) 25 T. The other physical conditions for the simulation are the same as in Figure 1. The integer numbers over the reference lines label the different plateaux.

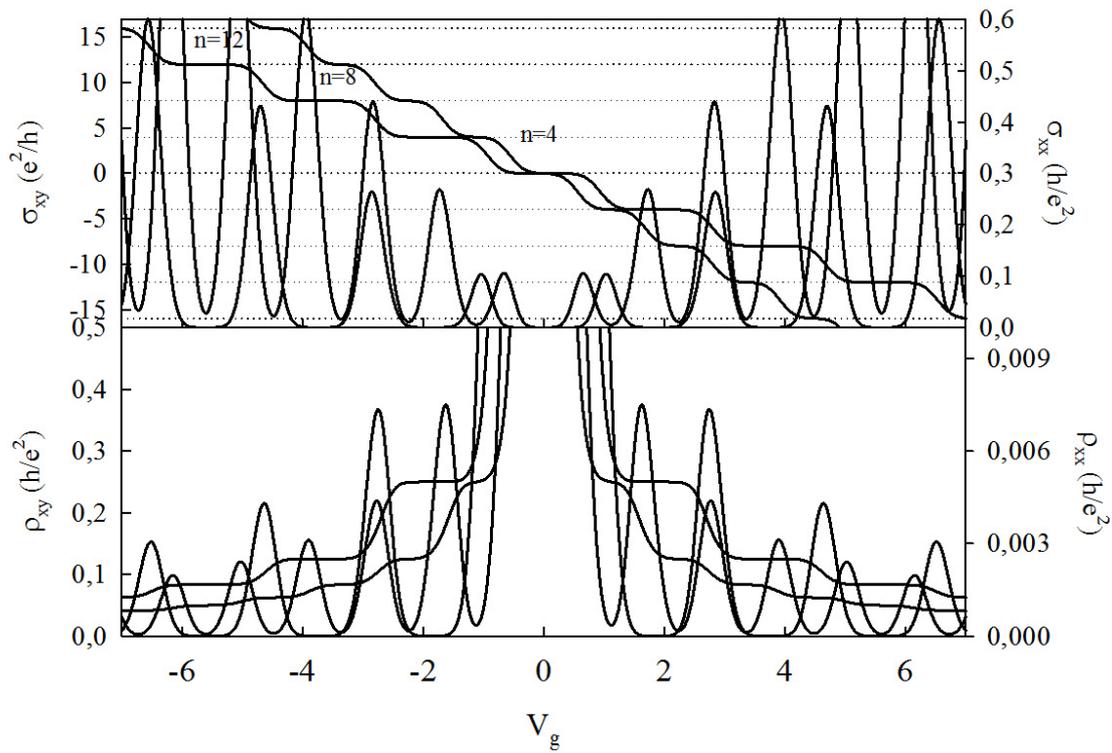

**Figure 3: Simulation with the model of the Hall magnetoconductivity, (a), and diagonal magnetoresistivities, (b), for the bilayer graphene as a function of the magnetic field**: To achieve the simulation we have considered a temperature of 1.6 K; an electron density $n_0=3\times10^{16}$ m$^{-2}$; a gyromagnetic factor of 2; and an effective mass of $m^*=0,05m_0$. Besides, we have take into account a constant Gaussian width with $\Gamma=0.025$ eV and a relaxation time $\tau=1$ ps. In figure (a) the integer numbers over the reference lines label the different plateaux.

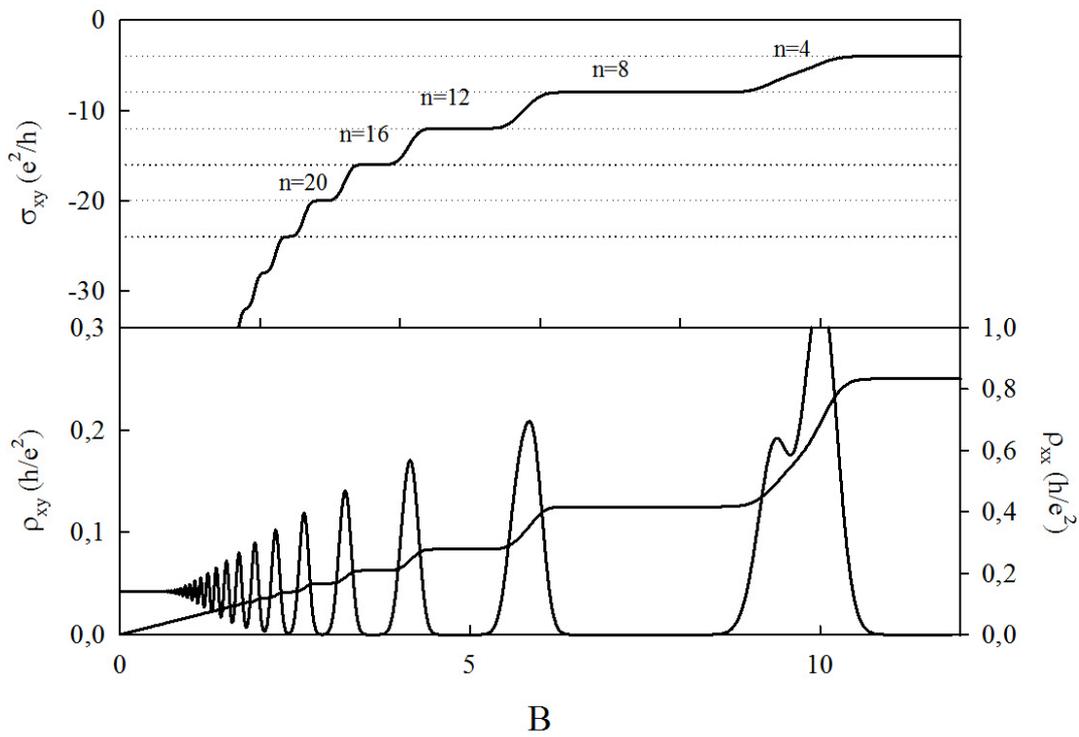